%% file: main.tex
\documentclass[10pt,twocolumn,letterpaper]{article}

\usepackage[pagenumbers]{iccv} 
\usepackage{xcolor}

\usepackage{multirow}
\usepackage{tcolorbox}
\definecolor{iccvblue}{rgb}{0.21,0.49,0.74}
\usepackage[pagebackref,breaklinks,colorlinks,allcolors=iccvblue]{hyperref}
\def\paperID{14} 
\def\confName{ICCV}
\def\confYear{2025}

\title{Prompting with Sign Parameters for Low-resource \\Sign Language Instruction Generation}

\author{Md. Tariquzzaman, Md Farhan Ishmam, Saiyma Sittul Muna, Md. Kamrul Hasan, Hasan Mahmud \\
Department of Computer Science and Engineering\\
Islamic University of Technology, Bangladesh\\
{\tt\small \{tariquzzaman,farhanishmam,saiymasittul,hasank,hasan\}@iut-dhaka.edu}
}

\begin{document}
\maketitle
\input{sec/0_abstract}
\input{sec/1_intro}

\input{sec/2_relatedwork}
\input{sec/3_bdsli}
\input{sec/4_results}
{
    \small
    \bibliography{main}
}

\end{document}

%% file: sec/0_abstract.tex
\begin{abstract}

Sign Language (SL) enables two-way communication for the deaf and hard-of-hearing community, yet many sign languages remain under-resourced in the AI space. Sign Language Instruction Generation (SLIG) produces step-by-step textual instructions that enable non-SL users to imitate and learn SL gestures, promoting two-way interaction. We introduce BdSLIG, the first Bengali SLIG dataset, used to evaluate Vision Language Models (VLMs) (i) on under-resourced SLIG tasks, and (ii) on long-tail visual concepts, as Bengali SL is unlikely to appear in the VLM pre-training data. To enhance zero-shot performance, we introduce Sign Parameter-Infused (SPI) prompting, which integrates standard SL parameters, like hand shape, motion, and orientation, directly into the textual prompts. Subsuming standard sign parameters into the prompt makes the instructions more structured and reproducible than free-form natural text from vanilla prompting. We envision that our work would promote inclusivity and advancement in SL learning systems for the under-resourced communities.

\end{abstract}

%% file: sec/1_intro.tex
\section{Introduction}
Sign Language (SL) differs fundamentally from spoken or written languages as it inherently requires communication in the visual modality. At a lexical level, these SL gestures are typically encapsulated in a short video of a few seconds \cite{bantupalli2018american, rayeed2023bdsl47}. With the current advancements in computer vision models \cite{dosovitskiy2020image,radford2021learning,videoTransformers}, there has been significant work on automated processing of SL video clips, primarily for American Sign Language (ASL) recognition tasks \cite{sharma2021asl,lee2021american}.

Despite the proliferation of ASL recognition works, we highlight two major research gaps in this domain. Firstly, the overwhelming focus on ASL works has marginalized non-ASL languages \cite{nc2022addressing}, limiting the \textit{inclusivity} and \textit{generalizability} of the models \cite{desai-etal-2024-systemic}. Secondly, SL recognition addresses only one half of the SL communication process, \ie, understanding the SL user. The other half, \ie, communicating back to the SL user, requires \textit{learning} and \textit{producing} the SL gestures - a task which is relatively less popular. 

Current literature has provided methods for non-ASL and under-resourced SL recognition tasks \cite{nc2022addressing,charuka2023sign}. For learning SL, LLM-based automated SL Instruction Generation (SLIG) \cite{inan2024generating} creates textual instructions that guide the users to replicate the SL gesture. Automated gesture synthesis \cite{fang2024signllm} can also be a plausible option for communicating back to the SL user, albeit with limited real-life applicability. However, there has been no prior work on the intersection of the aforementioned research problems, \ie, SLIG for low-resource languages. Our work serves as a pilot project to address this problem space.

SLIG in non-ASL languages is critical, as each SL possesses unique gestural cues, shaped by linguistic, cultural, and regional factors \cite{lucas2001importance,lutzenberger2023social}. These variations mirror linguistic differences in spoken or written languages. We chose Bengali for our non-ASL evaluation language as it remains an under-resourced language in the automated SL space \cite{rubaiyeat2024bdslw60}, while being native to 300 million active speakers, making it a high-impact choice for real-world applicability. Bengali SL also predominantly focuses on recognition tasks \cite{rayeed2023bdsl47,raihan2024bengali,miah2022bensignnet}; hence, our proposed BdSLIG is the first low-resource and Bengali SLIG dataset.


We hypothesize that a Vision Language Model (VLM) \cite{zhang2024vision,ishmam2024image}, when given a frame or short video of a sign, can generate a textual description (\textit{e.g.}, ``both hands raised near face, wrists bent inwards”) that captures the sign’s meaning and form. Such descriptions could serve as interactive instructions for learners (analogous to language learning feedback \cite{nakata2025student}) and also provide interpretability by explaining what the model ``sees.” This idea builds on prior work in explainable sign technology, \eg, the Learn2Sign system \cite{learn2signxai}, which provided feedback on location, movement, and handshape to ASL learners. Since Bengali Sign Language is a low-resource language and sign recognition models are still underdeveloped, it is essential to analyze how large models behave in this context. Prior research \cite{inan-etal-2025-signalignlm} has shown that general LLMs may produce overly generic or semantically off-base sign descriptions without domain-specific guidance. In light of these limitations, our contributions can be summarized as follows:

\begin{itemize}
\item We propose the first dataset to evaluate the performance of VLMs on the under-resourced task of Bengali sign language instruction generation.
\item We introduce Sign Parameter-Infused (SPI) prompting, which integrates standard sign language parameters, \eg, hand shape, motion, and orientation, directly into the textual prompts of the VLMs. Experiments reveal SPI prompting generally leads to better performance.
\item Our experiments also serve as an evaluation of the performance of VLMs on long-tail distributed and semantically fine-grained visual concepts.
\end{itemize}

%% file: sec/3_bdsli.tex
\section{Methodology}
In this section, we first define the sign language parameters (\S\ref{subsec:parameters}) that will be relevant to our prompting technique (\S\ref{subsec:promptingFramework}). We also construct the BdSLIG dataset (\S\ref{subsec:bdslig}) and define our evaluation framework (\S\ref{subsec:evaluation}).

\subsection{Sign Language Parameters}
\label{subsec:parameters}

Sign language parameters are essential visual cues used to identify or describe a sign language gesture \cite{sandler2006signParameters}. Our work uses the following parameter categories, defined below with a few associated keyword tags:

\begin{enumerate}
    \item \textbf{Handshape}: The morphological configuration, \eg, raised, extended, pointed, bent, spread, twisted, of the fingers and palms. 
    \item \textbf{Movement Type}: The trajectory or action of the hands, \eg, static, circular, squeezing, grabbing, rotating, pulling.
    \item \textbf{Location}: The placement of the hands while producing the sign, \eg, near head, forehead, eye, ear, nose, mouth, chest, shoulder.
    \item \textbf{Palm Orientation}: The direction the palm is facing relative to the signer's body, \eg, inward, outward, upward, downward, left, right.
    \item \textbf{Spatial Interaction}: The role of one or both hands in the sign's spatial scheme, \eg, right-hand active, left-hand active, both-hands symmetric, both-hands asymmetric.
    \item \textbf{Temporal Dynamics}: The temporal pattern of the movement onset, duration, and repetition, \eg, sustained hold, repeated oscillation, single tap, repeated tapping.
    \item \textbf{Facial Cues}: Facial expressions, \eg, neutral, brows-raised, brows-furrowed, mouth-open, mouth-closed, lip-pursed, that can disambiguate the sign language.
\end{enumerate}

\begin{figure*}[t]
    \centering
    \includegraphics[width=0.32\textwidth]{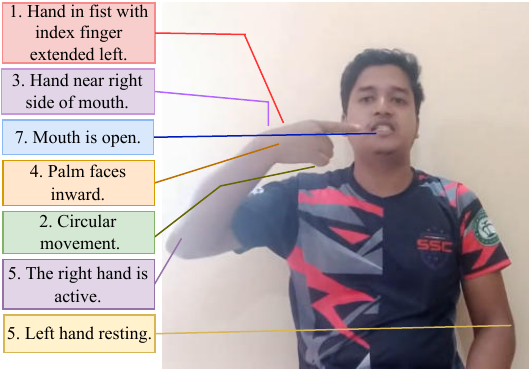}
    \hfill
    \includegraphics[width=0.32\textwidth]{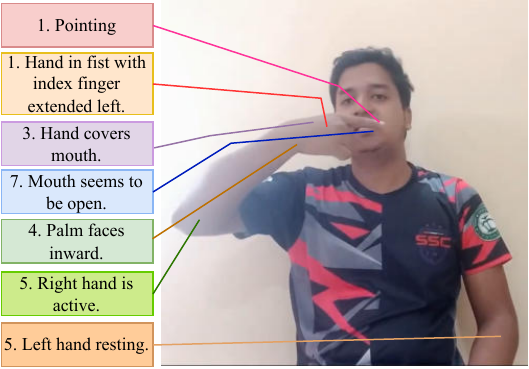}
    \hfill
    \includegraphics[width=0.32\textwidth]{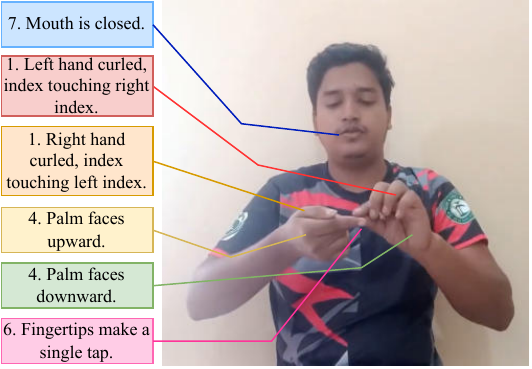}
    \caption{Parameter-based description of the \texttt{Toothpaste} sign (video taken from the BdSLIG dataset). The numbers denote the parameter numbers mentioned in \S\ref{subsec:parameters}. The parameter descriptions are given to elucidate and exemplify the given SL parameters and \textit{are not} part of the BdSLIG dataset.}
    \label{fig:sign-params-fig}
\end{figure*}

These sign-language parameters serve two roles in our work: (i) to establish the annotation guidelines for human annotation of the BdSLIG dataset, and (ii) to define the Sign Parameter-Infused (SPI) prompting for evaluating zero-shot capabilities of vision-language models (VLMs). \cref{fig:sign-params-fig} provides a visual overview of these parameters, where each category is illustrated with example video frames sampled from the BdSLIG dataset. 

\subsection{BdSLIG Dataset}
\label{subsec:bdslig}

The images of BdSLIG are sourced from the BdSLW60 dataset~\cite{rubaiyeat2024bdslw60}, a 60-word Bengali sign language recognition dataset. For each word, we selected one representative video and manually annotated the textual sign instructions. We hired three trained annotators following the annotation guidelines with our defined sign parameters (\S\ref{subsec:parameters}). The annotations were validated in two stages: first for grammatical and semantic consistency of the text, then by a Bengali sign language domain expert. Both the annotators and validators received monetary compensation for their work.


\subsection{Prompting Framework}
\label{subsec:promptingFramework}
Our prompting framework consists of three stages: (i) Frame Sampling, (ii) Sign Parameter Injection, and (iii) Instruction Generation. The second stage uses Sign Parameter-Infused (SPI) prompting, whereas the corresponding vanilla prompting strategy disregards this step.


\subsubsection{Frame Sampling}
\label{subsec:frameSampling}
A video input is taken from the framerate-standardized version, where all videos are adjusted to 30 frames per second. As inferring a high number of frames can be challenging to the VLM and increase the computational overhead, we uniformly sample every $20^{th}$ frame from the video. This reduces the input length and thereby decreases the context fed to the VLM while retaining most of the temporal information.

The limitations of uniform sampling of video frames in the context of sign languages are two-fold. Firstly, the motion of the sign is not uniform, \ie, there can be periods of high and low activities. Hence, extracting an equal number of frames from high and low activities should not be optimal. Secondly, every $k^{th}$ frame sampled uniformly may not be a \textit{key frame}. We define the key frames as the minimum set of frames that capture the motion of the sign language. While there can be subjectivity in picking the key frames, the extraction procedure will surely not be uniform. Despite these limitations, we decided to adopt uniform sampling for its simplicity, and as our work serves as a pilot study.

\subsubsection{Sign Parameter Injection}
\label{subsec:signParamInjection}
Given a set of sign language parameters $\mathcal{S}$ and a base textual prompt $\mathcal{P}$, we define our Sign Parameter-Infused (SPI) prompt $f(\mathcal{P},\mathcal{S})$. Specifically, our SPI prompting describes the visual motor cues, as seen in \cref{fig:sign-params-fig}, to obtain a stepwise, narrative-style instruction set.


\subsubsection{Instruction Generation}
\label{subsec:instructionGeneration}
We define the generated instruction as a sequence of answer tokens, $\mathbf{a}=\langle a_1,a_2,\dots,a_n\rangle$. For a generative Vision Language Model (VLM), $g$, taking a video feed or a set of video frames $\mathcal{V}$ and a prompt $\mathcal{P}$, we define instruction generation for vanilla prompting:
\begin{equation}
   a_i = \arg\max\mathbb{P}(a|s(\mathcal{V}),\mathcal{P},a_{<i}), 
\end{equation}
where $s(\cdot)$ is the video sampling function. Similarly, for SPI prompting, the generated instructions are:
\begin{equation}
    a_i = \arg\max\mathbb{P}(a|s(\mathcal{V}),f(\mathcal{P},\mathcal{S}),a_{<i}), 
\end{equation}
where $\mathcal{S}$ is the set of sign language parameters.


\begin{table*}[t]
\centering
\small
\setlength{\tabcolsep}{5pt} 
\renewcommand{\arraystretch}{0.95} 

\begin{tabular}{lccccccc} 
\toprule
\multirow{2}{*}{\textbf{Prompting}} & \multirow{2}{*}{\textbf{Model}} & \multicolumn{6}{c}{\textbf{Metric}} \\
\cmidrule{3-8}
 &  & \textbf{ROUGE-1} & \textbf{ROUGE-2} & \textbf{ROUGE-L} & \textbf{BLEU} & \textbf{METEOR} & \textbf{BERTScore}\\
\midrule
\multirow{4}{*}{Vanilla} 
 & GPT 4.1-mini & 0.526  &  0.232  &  0.342 & 0.174 &  0.396 & 0.396 \\
 & GPT 4.1      & 0.528  &  0.220 &  0.335 & 0.149 &  0.416 &  0.394 \\
 & Gemini 2.0 Flash & 0.464  &  0.209  &  0.328 & 0.141 &  0.329  &  0.416 \\
 & Gemini 2.5 Pro   & 0.505  &  0.224  &  0.338 & 0.166 &  0.354  &  0.396 \\
\midrule
\multirow{4}{*}{SPI (Ours)} 
 & GPT 4.1-mini & 0.522 \textcolor{red}{↓}  &  0.224 \textcolor{red}{↓}  &  0.326 \textcolor{red}{↓} & 0.162 \textcolor{red}{↓} & 0.447 \textcolor{green}{↑}   &     0.396 {\textcolor{gray}{=}} \\
 & GPT 4.1      &  \textbf{0.553} \textcolor{green}{↑} &   \textbf{0.271} \textcolor{green}{↑} &  \textbf{0.384} \textcolor{green}{↑} & \textbf{0.196} \textcolor{green}{↑} &  \textbf{0.492}   \textcolor{green}{↑}      & 0.441 \textcolor{green}{↑} \\
 & Gemini 2.0 Flash & 0.505 \textcolor{green}{↑} &  0.250 \textcolor{green}{↑} &  0.365 \textcolor{green}{↑} & 0.168 \textcolor{green}{↑} &  0.359 \textcolor{green}{↑}  &      \textbf{0.456} \textcolor{green}{↑} \\
 & Gemini 2.5 Pro   &  0.507 \textcolor{green}{↑} &  0.212 \textcolor{red}{↓}  &  0.319 \textcolor{red}{↓} & 0.141 \textcolor{red}{↓} &  0.365  \textcolor{green}{↑}  &     0.387 \textcolor{red}{↓} \\
\bottomrule
\end{tabular}
\caption{Evaluation of Vanilla and SPI prompting on our BdSLIG dataset.}
\label{tab:eval_metrics_acl}
\end{table*}
\subsection{Evaluation Framework}
\label{subsec:evaluation}
We primarily have two evaluation goals: (i) assessing the quality of the generated instructions on the BdSLIG dataset and (ii) evaluating the efficacy of SPI prompting over vanilla prompting. A quantitative assessment can be performed by benchmarking state-of-the-art VLMs (\S\ref{subsubsec:Model}) and comparing vanilla and SPI prompting strategies using standard text evaluation metrics (\S\ref{subsubsec:metrics}).

\subsubsection{Models}
\label{subsubsec:Model}
We evaluate four state-of-the-art VLMs: (i) GPT 4.1-mini, (ii) GPT 4.1, (iii) Gemini 2.0 Flash, and (iv) Gemini 2.5 Pro. All these are proprietary models with reasonably high scores in LLM benchmarks, \eg, LMArena leaderboard\footnote{\url{https://lmarena.ai/leaderboard}}.

\subsubsection{Evaluation Metrics}
\label{subsubsec:metrics}
The evaluation of sign language instructions using functions, heuristics, and metrics can be tricky. We first emphasize that ideal evaluation can only be achieved by a human, preferably a sign language educator or expert. Text matching or similarity metrics \textit{do not} evaluate the correctness of the generated instruction. The metrics capture text-level and semantic similarity. The suite of metrics used in our benchmarks is: ROUGE-1, ROUGE-2, ROUGE-L, BLEU, METEOR, and BERTScore. Each metric has its own limitations, and we believe using all would provide a holistic view of the generated textual instructions.



%% file: sec/4_results.tex
\section{Results and Analysis}
This section is centered around our benchmark results from \cref{tab:eval_metrics_acl}. We also present several analyses and discussions on the validity of the results and scopes for improvement.
\subsection{SPI \vs Vanilla Prompting} \cref{tab:eval_metrics_acl} reports how well VLM-generated instructions align with our expert annotation from the BdSLIG dataset. SPI prompting consistently outperforms Vanilla Prompting across most metrics, especially METEOR and BERTScore. These metrics emphasize semantic similarity and are less sensitive to surface-level lexical variations, suggesting that SPI prompting helps the VLMs generate more structured and semantically faithful instructions, crucial in low-resource sign language contexts.

\subsection{Larger \vs Smaller Models}
Larger models like GPT-4.1 and Gemini-2.5-Pro benefit most, showing strong gains in both lexical and semantic alignment. While smaller variants like GPT-4.1-mini and Gemini 2.0 Flash also show improvement under the SPI setting, their limited capacity to interpret nuanced motion cues leads to weaker alignment with ground-truth annotations. Overall, GPT-4.1 takes the crown by achieving the highest score in all metrics except BERTScore.

\subsection{Can VLMs generate acceptable instructions?}
Answering this question can be quite nuanced. The general text scores, both lexical and semantic matching, can be deemed relatively low. If we look at other text generation tasks, \eg, machine translation, the generative models tend to perform much better. However, most such tasks have \textit{objective} ground truths, \ie, there is less flexibility in annotating the answer. Instructions required to perform or imitate a sign are very \textit{subjective}, \ie, there can be multiple ways to write the instructions to perform the same sign.

\subsection{How accurate are the metrics?}
Following \S\ref{subsubsec:metrics}, we emphasize that lexical and semantic similarity metrics are not suitable for evaluating sign language instructions. To elucidate, we present two first-step instructions for the sign word \texttt{toothpaste} (\cref{fig:sign-params-fig}).

\begin{tcolorbox}[colback=white,colframe=black,title=\textcolor{white}{Sample-1: Instruction Step-1 for \textit{toothpaste}},coltitle=white]
The right hand, with fingers gently curved and palm facing toward the signer, is brought up to the mouth area. The fingertips lightly touch or hover near the lips, with the hand at chin or mouth level. The left hand remains at rest.
\end{tcolorbox}

\begin{tcolorbox}[colback=white,colframe=black,title=\textcolor{white}{Sample-2: Instruction Step-1 for \textit{toothpaste}},coltitle=white]
The right hand moves toward the mouth with the index finger fully extended and pointing straight ahead, while the other fingers are curled into a loose fist. At the same time, the mouth opens slightly, revealing the teeth clearly.
\end{tcolorbox}
These two sample instructions essentially imitate the same hand motion. To a human learner, the differences in execution are small. Yet, due to the variations in wording, order of description, and emphasis on certain body positions, the text evaluation metrics would produce a considerably lower score. This is evident as most of these metrics quantify surface-level lexical overlap. Metrics like BertScore convey some form of semantic similarity, but not in the context of sign language. A possible approach is to use a larger and more contextual metric to evaluate the text pairs, \eg, using LLM-based evaluation methods \cite{li2024llms}.

\subsection{Long-tail Evaluation}
The current generation of VLMs is trained on large-scale web corpora \cite{radford2021learning}. For most evaluation tasks, \eg, question answering on natural images, it is highly likely that the evaluation data is present in the pre-training corpus of the VLM \cite{li2024task}. However, as Bengali is a low-resource language, it is highly unlikely that the VLMs are pre-trained on Bengali sign language and the associated instruction data, \ie, it is in the tail end of the pre-training corpus distribution. Hence, BdSLIG can be a good benchmark for visual long-tail distribution. Following this setting, GPT-4.1 takes the crown as the best-performing model in both prompting strategies as per \cref{tab:eval_metrics_acl}. Hence, GPT-4.1 should have relatively better visual understanding and grounding.

\section{Conclusion}
We present (i) BdSLIG, a novel Bengali sign language instruction generation dataset, and (ii) Sign Parameter-Infused (SPI) prompting, a prompting strategy that injects sign language parameters into vision-language prompts. Instruction generation following standard sign parameters makes the description structured and reproducible, unlike free-form text. Canonicalizing the descriptors can be suitable for classification, retrieval, or alignment tasks. Our experiments investigate the efficacy of SPI prompting, long-tail evaluation of state-of-the-art models, and metrics for evaluating sign language instruction. Our work serves as a pilot project to promote research in multimodal generative systems for individuals with accessibility needs.

\section*{Reproducibility}
Our dataset is available on HuggingFace: \url{https://huggingface.co/datasets/aplycaebous/BdSLIG} and the code is available on GitHub: \url{https://github.com/tariquzzamanf/SPIP}. The reproducibility of results on certain proprietary models can vary based on the time and date of model inference.

